\documentclass[10pt]{article}
\usepackage{graphicx}

\def\Title#1{\begin{center} {\Large #1 } \end{center}}
\def\Author#1{\begin{center}{ \sc #1} \end{center}}
\def\Address#1{\begin{center}{ \it #1} \end{center}}

\newcommand\pubblock{\rightline{\begin{tabular}{l} Proceedings of the Fifth Annual LHCP\\ \pubnumber\\
         \pubdate  \end{tabular}}}

\newenvironment{Abstract}{\begin{quotation} \begin{center} 
             \large ABSTRACT \end{center}\bigskip 
      \begin{center}\begin{large}}{\end{large}\end{center} \end{quotation}}

\newenvironment{Presented}{\begin{quotation} \begin{center} 
             PRESENTED AT\end{center}\bigskip 
      \begin{center}\begin{large}}{\end{large}\end{center} \end{quotation}}

\def\beq{\begin{equation}}
\def\eeq#1{\label{#1}\end{equation}}
\def\eeqn{\end{equation}}

\def\beqa{\begin{eqnarray}}
\def\eeqa#1{\label{#1}\end{eqnarray}}
\def\eeqan{\end{eqnarray}}

\let\bar=\overbar

\def\Dslash{\not{\hbox{\kern-4pt $D$}}}
\def\dslash{\not{\hbox{\kern-2pt $\del$}}}

\def\msb{{\bar{\ssstyle M \kern -1pt S}}}

\usepackage{amsmath,amssymb,amsfonts,cancel,bbm}
\usepackage{array,booktabs,colortbl,multirow}
\usepackage{color}
\usepackage[dvipsnames]{xcolor}
\usepackage{rotating}
\usepackage{cite}
\usepackage{url}
\usepackage{floatrow}

\newfloatcommand{capbtabbox}{table}[][\FBwidth]

\newcommand{\units}[1]{~\mathrm{#1}}

\newcommand{\cmrule}{\midrule[0.25mm]}

\newcommand\pubnumber{ }

\newcommand\pubdate{October 15, 2017}

\begin{document}

\large
\begin{titlepage}
\pubblock

\vfill
\Title{The Global Electroweak and Higgs Fits in the LHC era}
\vfill


\renewcommand{\thefootnote}{\fnsymbol{footnote}}
\setcounter{footnote}{1}
        
\Author{Jorge de Blas$^a$\footnote{Speaker}, Marco Ciuchini$^b$, Enrico Franco$^c$, Satoshi Mishima$^d$,\\ Maurizio Pierini$^e$, Laura Reina$^f$ and Luca Silvestrini$^c$}
\Address{$^a$ Universit\`{a} di Padova and INFN, Sezione di Padova, Via Marzolo 8, I-35131 Padova, Italy}
\Address{$^b$INFN, Sezione di Roma Tre, Via della Vasca Navale 84, I-00146 Roma, Italy}
\Address{$^c$INFN, Sezione di Roma, Piazzale A. Moro 2, I-00185 Rome, Italy}
\Address{$^d$Theory Center, IPNS, KEK, 1-1 Oho, Tsukuba, 305-0801, Japan}
\Address{$^e$CERN, Geneva, Switzerland}
\Address{$^f$Physics Department, Florida State University, Tallahassee, FL 32306-4350, USA}

\vfill
\begin{Abstract}

We update the global fit to electroweak precision observables, including the effect of the latest measurements at hadron colliders of the $W$ and top-quark masses and the effective leptonic weak mixing angle. 
We comment on the impact of these measurements in terms of constraints on new physics.
We also update the bounds derived from the fit to the Higgs-boson signal strengths, including the observables measured at the LHC Run 2, and compare the improvements with respect to the 7 and 8 TeV results.

\end{Abstract}
\vfill

\begin{Presented}
The Fifth Annual Conference\\
 on Large Hadron Collider Physics \\
Shanghai Jiao Tong University, Shanghai, China\\ 
May 15-20, 2017
\end{Presented}
\vfill
\end{titlepage}
\def\thefootnote{\fnsymbol{footnote}}
\setcounter{footnote}{0}
%

\normalsize 


\renewcommand{\thefootnote}{\arabic{footnote}}
\setcounter{footnote}{0}



\section{Introduction}

Recently, the ATLAS collaboration presented the first LHC measurement of the $W$ mass ($M_W$)~\cite{Aaboud:2017svj}, with a precision comparable to the LEP2 and Tevatron combination. With this, the LHC experiments keep pushing the frontiers of our knowledge of the electroweak scale, not only via direct searches but also through indirect tests of new physics (NP) in electroweak precision observables (EWPO). Similarly, the latest results of the effective leptonic weak mixing angle ($\sin^2{\theta_{\mathrm{eff}}^{\mathrm{lept}}}$) at the Tevatron~\cite{CDF:2016cry} confirmed the potential of hadron colliders (HC) for precision measurements beyond the $W$ mass. 
It is therefore interesting to study the impact of these recent measurements within the context of the global electroweak fit and in setting constraints on physics beyond the Standard Model (SM).

Turning our attention to the Higgs boson, the negative evidence from the LHC Run 1 of any NP effects on the Higgs signal strengths at the $\sim 10\%$ level is corroborated by the latest data at 13 TeV. As these Run-2 results become comparable in precision with (in some cases more precise than) the 7/8 TeV ones, an updated combination is timely in order to asses the constraining power of the the full LHC Higgs data set.

In this proceedings we present the latest updates in the electroweak and Higgs boson observable fits. 
Section \ref{sec:EWfit} covers the study of the fit to EWPO, while the status of the fit to Higgs boson signal strengths is discussed in Section \ref{sec:Higgsfit}. A more extended study, discussing also the interplay between both types of constraints,  
will be presented in a separate publication. 
All the fits presented in this note have been performed using the {\tt HEPfit} code \cite{HEPfit}. 


\section{The global electroweak fit and precision observables at the LHC}
\label{sec:EWfit}

Compared to our previous fits presented in \cite{deBlas:2016nqo,deBlas:2016ojx}, 
in these proceedings we include the following updates in the experimental measurements taken at HC: 
1) The 2016 determinations of the top-quark mass ($m_t$) from the Tevatron and LHC experiments.
Each of these measurements exceeds individually the precision of the previous world average from 2014. It must be noted though that, currently, only the individual measurements of $m_t$ from each experiment are available. Their uncertainties are however expected to have a significant correlation and, therefore, a weighted average of these measurements may not be appropriate. Moreover, the independent determinations from CMS and the Tevatron experiments differ by more than $1~\!\sigma$, so any combination must be interpreted carefully. As a first approximation, however, one can still perform such weighted average rescaling the error according to the method in \cite{Patrignani:2016xqp}, and use $m_t=173.1\pm 0.6\units{GeV}$.\footnote{We also consider an extra 0.5 GeV error associated with the interpretation of the experimental Monte Carlo mass as the top-quark pole mass.} 
2) The $W$ mass measurement from ATLAS. As in the case of $m_t$, the uncertainty on the $M_W$ determinations from ATLAS and the Tevatron experiments are expected to have some correlated components. In this case, however, we observe that, assuming a not too large source of common uncertainty between both measurements, the $W$ mass average is relatively stable. We take $M_W=80.379\pm 0.012\units{GeV}$. 3) The Tevatron combination and LHC measurements of $\sin^2{\theta_{\mathrm{eff}}^{\mathrm{lept}}}$. In this case the measurements agree reasonably well and, 
to study their impact on the fit in first approximation,
we simply take a normal weighted average.
Apart from all these HC measurements, we also updated the fit with the latest determination of $\alpha_S(M_Z)$ \cite{Patrignani:2016xqp}. 
Finally, on the theory side, the calculation of the bottom asymmetries has been updated with the 2-loop bosonic contributions to $\sin{\theta_{\rm eff}^b}$ from \cite{Dubovyk:2016aqv}.
\begin{table}[t]
{\footnotesize
\begin{center}
\begin{tabular}{lcccc}
\toprule
& Measurement & Posterior & Prediction &Pull \\
\cmrule
$\alpha_{s}(M_Z)$ & $ 0.1180 \pm 0.0010 $ & $ 0.1180 \pm 0.0009 $  & $ 0.1184 \pm 0.0028 $ & -0.1  \\ 
$\Delta\alpha^{(5)}_{\rm had}(M_Z)$ & $ 0.02750 \pm 0.00033 $ & $ 0.02743 \pm 0.00025 $  & $ 0.02734 \pm 0.00037 $ & \phantom{-}0.3  \\ 
$M_Z$ [GeV] & $ 91.1875 \pm 0.0021 $ & $ 91.1880 \pm 0.0021 $  & $ 91.198 \pm 0.010 $ & -1.0  \\ 
$m_t$ [GeV] & $ 173.1 \pm 0.6 \pm 0.5 $ & $ 173.43 \pm 0.74 $  & $ 176.1 \pm 2.2 $ & -1.3  \\ 
$m_H$ [GeV] & $ 125.09 \pm 0.24 $ & $ 125.09 \pm 0.24 $  & $ 100.6 \pm 23.6 $ & \phantom{-}1.0  \\ 
\cmrule
$M_W$ [GeV] & $ 80.379 \pm 0.012 $ & $ 80.3643 \pm 0.0058 $  & $ 80.3597 \pm 0.0067 $ & \phantom{-}1.4  \\ 
$\Gamma_{W}$ [GeV] & $ 2.085 \pm 0.042 $ & $ 2.08873 \pm 0.00059 $  & $ 2.08873 \pm 0.00059 $ & -0.1  \\ 
$\sin^2\theta_{\rm eff}^{\rm lept}(Q_{\rm FB}^{\rm had})$ & $ 0.2324 \pm 0.0012 $ & $ 0.231454 \pm 0.000084 $  & $ 0.231449 \pm 0.000085 $ & \phantom{-}0.8  \\ 
$P_{\tau}^{\rm pol}={A}_\ell$ & $ 0.1465 \pm 0.0033 $ & $ 0.14756 \pm 0.00066 $  & $ 0.14761 \pm 0.00067 $ & -0.3  \\ 
$\Gamma_{Z}$ [GeV] & $ 2.4952 \pm 0.0023 $ & $ 2.49424 \pm 0.00056 $  & $ 2.49412 \pm 0.00059 $ & \phantom{-}0.5  \\ 
$\sigma_{h}^{0}$ [nb] & $ 41.540 \pm 0.037 $ & $ 41.4898 \pm 0.0050 $  & $ 41.4904 \pm 0.0053 $ & \phantom{-}1.3  \\ 
$R^{0}_{\ell}$ & $ 20.767 \pm 0.025 $ & $ 20.7492 \pm 0.0060 $  & $ 20.7482 \pm 0.0064 $ & \phantom{-}0.7  \\ 
$A_{\rm FB}^{0, \ell}$ & $ 0.0171 \pm 0.0010 $ & $ 0.01633 \pm 0.00015 $  & $ 0.01630 \pm 0.00015 $ & \phantom{-}0.8  \\ 
${A}_{\ell}$ (SLD) & $ 0.1513 \pm 0.0021 $ & $ 0.14756 \pm 0.00066 $  & $ 0.14774 \pm 0.00074 $ & \phantom{-}1.6  \\ 
$R^{0}_{b}$ & $ 0.21629 \pm 0.00066 $ & $ 0.215795 \pm 0.000027 $  & $ 0.215793 \pm 0.000027 $ & \phantom{-}0.7  \\  
$R^{0}_{c}$ & $ 0.1721 \pm 0.0030 $ & $ 0.172228 \pm 0.000020 $  & $ 0.172229 \pm 0.000021 $ & -0.05  \\ 
$A_{\rm FB}^{0, b}$ & $ 0.0992 \pm 0.0016 $ & $ 0.10345 \pm 0.00047 $  & $ 0.10358 \pm 0.00052 $ & -2.6  \\ 
$A_{\rm FB}^{0, c}$ & $ 0.0707 \pm 0.0035 $ & $ 0.07394 \pm 0.00036 $  & $ 0.07404 \pm 0.00040 $ & -0.9  \\ 
${A}_b$ & $ 0.923 \pm 0.020 $ & $ 0.934787 \pm 0.000054 $  & $ 0.934802 \pm 0.000061 $ & -0.6  \\ 
${A}_c$ & $ 0.670 \pm 0.027 $ & $ 0.66813 \pm 0.00029 $  & $ 0.66821 \pm 0.00032 $ & \phantom{-}0.1  \\ 
$\sin^2\theta_{\rm eff}^{\rm lept}(\mbox{Tev/LHC})$ & $ 0.23166 \pm 0.00032 $ & $ 0.231454 \pm 0.000084 $  & $ 0.231438 \pm 0.000087 $ & \phantom{-}0.7  \\
\bottomrule
\end{tabular}
\end{center}
}
\vspace{-0.25cm}
\caption{
  Experimental measurement, posterior, prediction, and pull for the 5 input
  parameters ($\alpha_s(M_Z)$, $\Delta \alpha^{(5)}_{\mathrm{had}}(M_Z)$, $M_Z$,
  $m_t$, $m_H$), and for the main EWPO considered in the SM fit. The values in
  the column \emph{Prediction} are determined without using the
  experimental information for the corresponding observable.} 
\label{tab:SMfit}
\end{table}
Using the preliminary combinations detailed in the previous lines we present in this section an (equally preliminary) update of the electroweak fit.\footnote{The HC determinations of $\sin{\theta_{\rm eff}^{\mathrm{lept}}}$ were already taken into account in the fit in \cite{deBlas:2016ojx}, but the average included here uses the Tevatron combination instead of the independent CDF and D0 measurements.} 
A more detailed study will be presented when more reliable combinations taking into account all correlated effects are provided by the corresponding experimental groups. 

\begin{figure}[h]
\begin{floatrow}
\capbtabbox{
\centering
{\small
\begin{tabular}{c c rrr}
 \toprule
 & ~~Result & \multicolumn{3}{c}{Correlation Matrix} \\ 
 \cmrule
\!\!\!\!$S$ & $\phantom{+} 0.09 \pm 0.10 $ & $\phantom{(}1.00\phantom{)}$ &&\!\!\!\!\!\!\\ 
\!\!\!\!$  $ & $\phantom{+} (0.08 \pm 0.10)$ &&&\!\!\!\!\!\!\\ 
\!\!\!\!$T$ & $\phantom{+} 0.11 \pm 0.12 $ & $\phantom{(}0.86\phantom{)}$ & $\phantom{(}1.00\phantom{)}$&\!\!\!\!\!\!\\ 
\!\!\!\!$  $ & $\phantom{+} (0.11 \pm 0.12)$ & $(0.85)$&&\!\!\!\!\!\!\\
\!\!\!\!$U$ & $-0.01 \pm 0.09 $ & $\phantom{(}-0.56\phantom{)}$ & $\phantom{(}-0.84\phantom{)}$ & $\phantom{(}1.00\phantom{)}$\!\!\!\!\!\! \\
\!\!\!\!$  $ & $\phantom{+} (0.00 \pm 0.09)$ &$(-0.49)$ & $(-0.79)$&\!\!\!\!\!\!\\
\bottomrule
 \toprule
\!\!\!\!$S$ & $\phantom{+} 0.09 \pm 0.08 $ & $\phantom{(}1.00\phantom{)}$ &\!\!\!\!\!\!\\ 
\!\!\!\!$  $ & $\phantom{+} (0.08 \pm 0.09)$ &&&\!\!\!\!\!\!\\ 
\!\!\!\!$T$ & $\phantom{+} 0.10 \pm 0.06 $ & $\phantom{(}0.87\phantom{)}$ & $\phantom{(}1.00\phantom{)}$ &\!\!\!\!\!\!\\ 
\!\!\!\!$  $ & $\phantom{+} (0.11 \pm 0.07)$ &$(0.86)$&&\!\!\!\!\!\!\\ 
\!\!\!\!$(U=0)$ &  &  &  &\!\!\!  \\
\bottomrule
 \end{tabular}
 }
 }{
\caption{Results of the fit for the oblique parameters $S$, $T$, $U$; and $S$, $T$ $(U=0)$. Results without the updates from HC are given in parenthesis.}
\label{tab:STUfit}
}
\hspace{-0.2cm}
\ffigbox{
\centering
  \hspace{-0.1cm}\includegraphics[width=.45\textwidth]{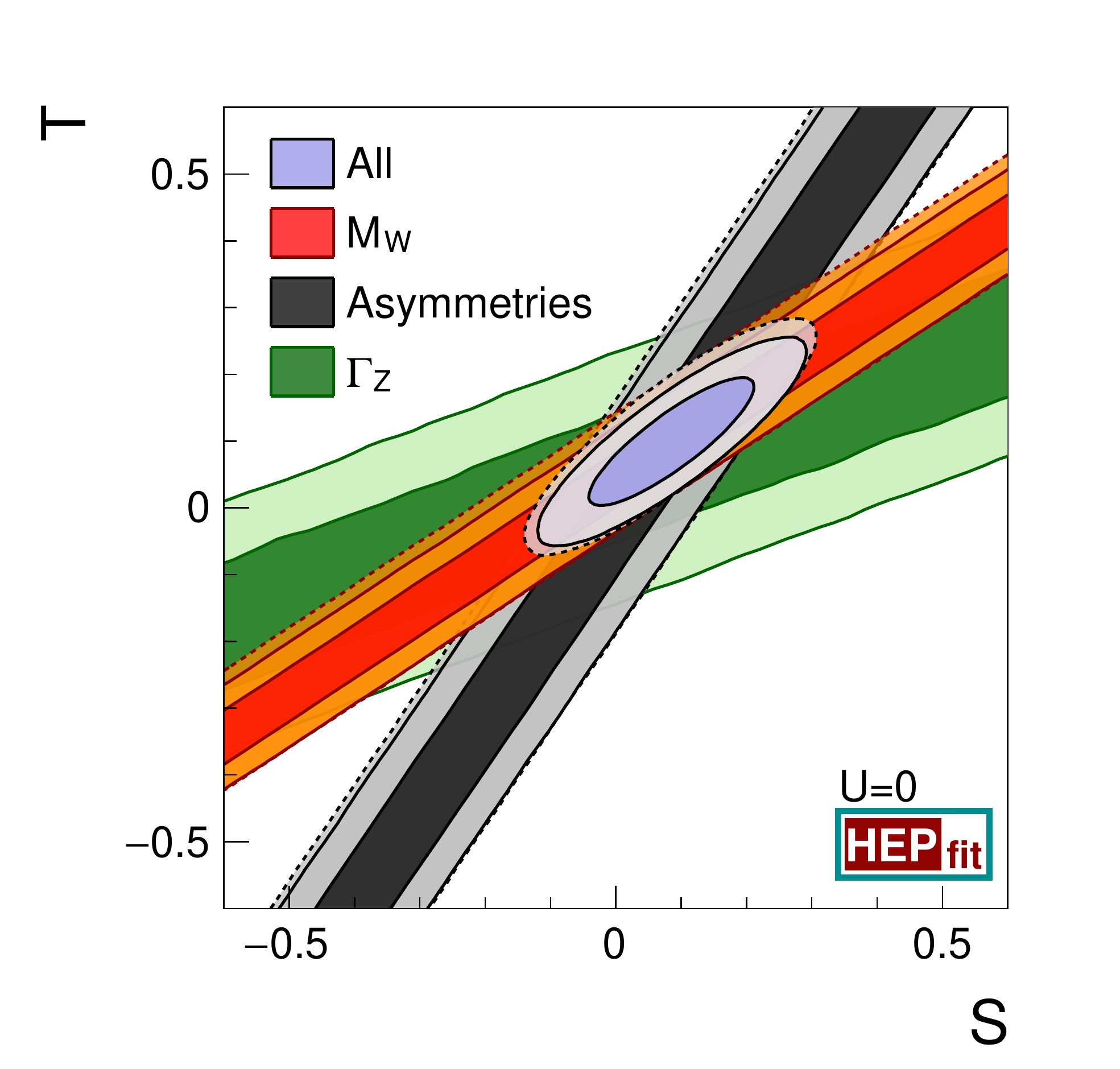}
  \vspace{-0.3cm}
}{
  \caption{ $68\%$ and $95\%$ probability contours for $S$ and $T$
    $(U=0)$, together with the individual constraints from $M_W$,
    the asymmetry parameters $\sin^2\theta_{\rm eff}^{\rm lept}$,
    $P_\tau^{\rm pol}$, $A_f$, and $A_{\rm FB}^{0,f}$ $(f=\ell,c,b)$, 
    and $\Gamma_Z$. Dashed lines indicate the results from the fit 
    without the updates from HC EWPO.~\label{fig:ST}}
}
\end{floatrow}
\end{figure}

The results of the updated SM fit are detailed in Table~\ref{tab:SMfit}. As expected, given that the SM fit is already overconstrained, the $m_t$-induced
parametric uncertainties were already subdominant compared to the experimental errors, and the updates are consistent with previous determinations, 
the effect of the HC updates on the SM fit is minimal.
Table~\ref{tab:STUfit} and Figure~\ref{fig:ST}, on the other hand, show the results for the fit to the oblique parameters $S$, $T$ and $U$, which are expected to be more sensitive to the updated observables. 
Small changes in the output of the $ST$ fit ($U=0$) can be observed at the 10$\%$ level. The role of each of the updated measurements in this small changes is summarized in Figure~\ref{fig:ST}.

\begin{figure}[h!]
\begin{center}
  \begin{tabular}{c c}
 \hspace{-1.1cm}\includegraphics[width=.54\textwidth]{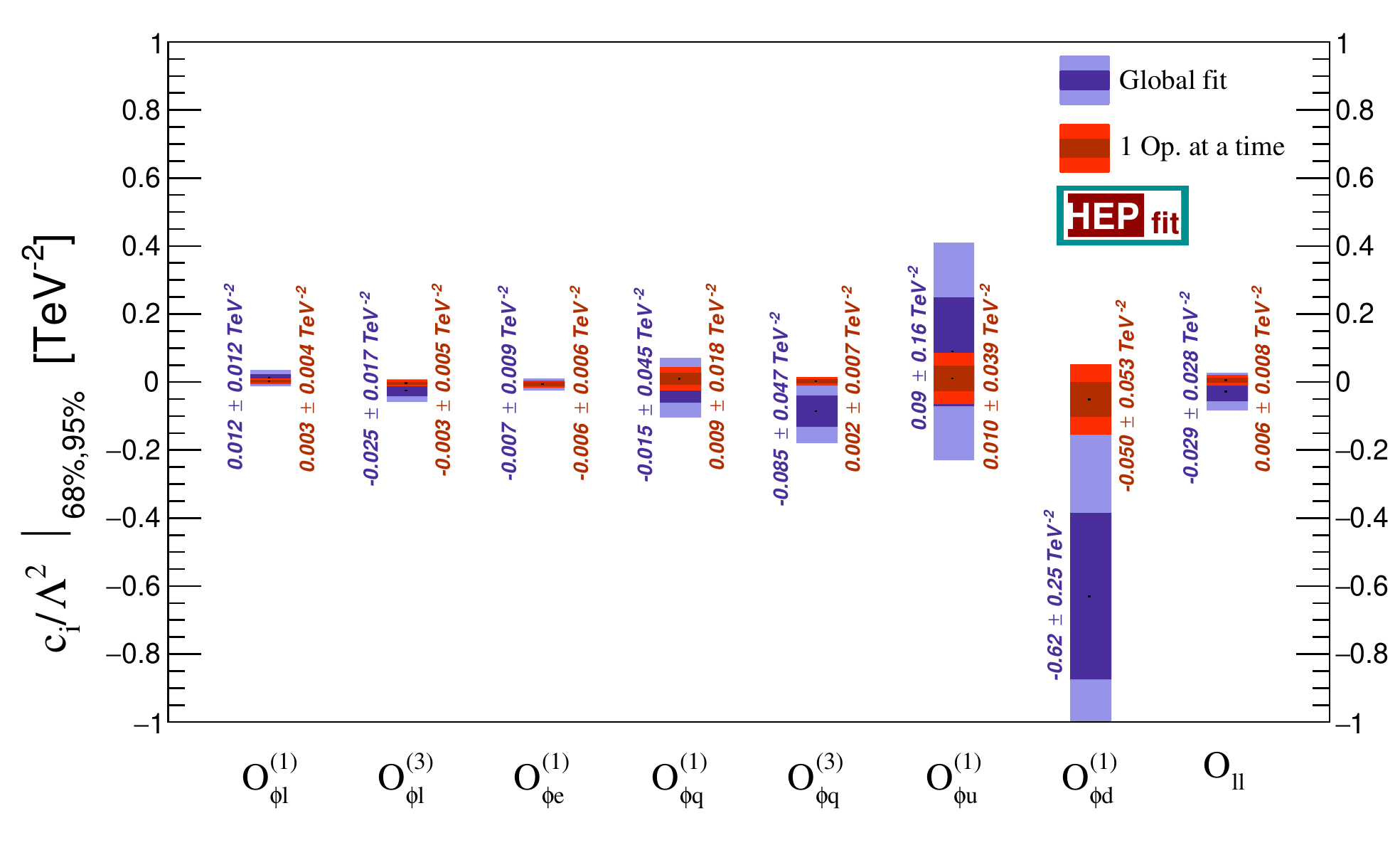} 
&
  \hspace{-0.4cm}\includegraphics[width=.54\textwidth]{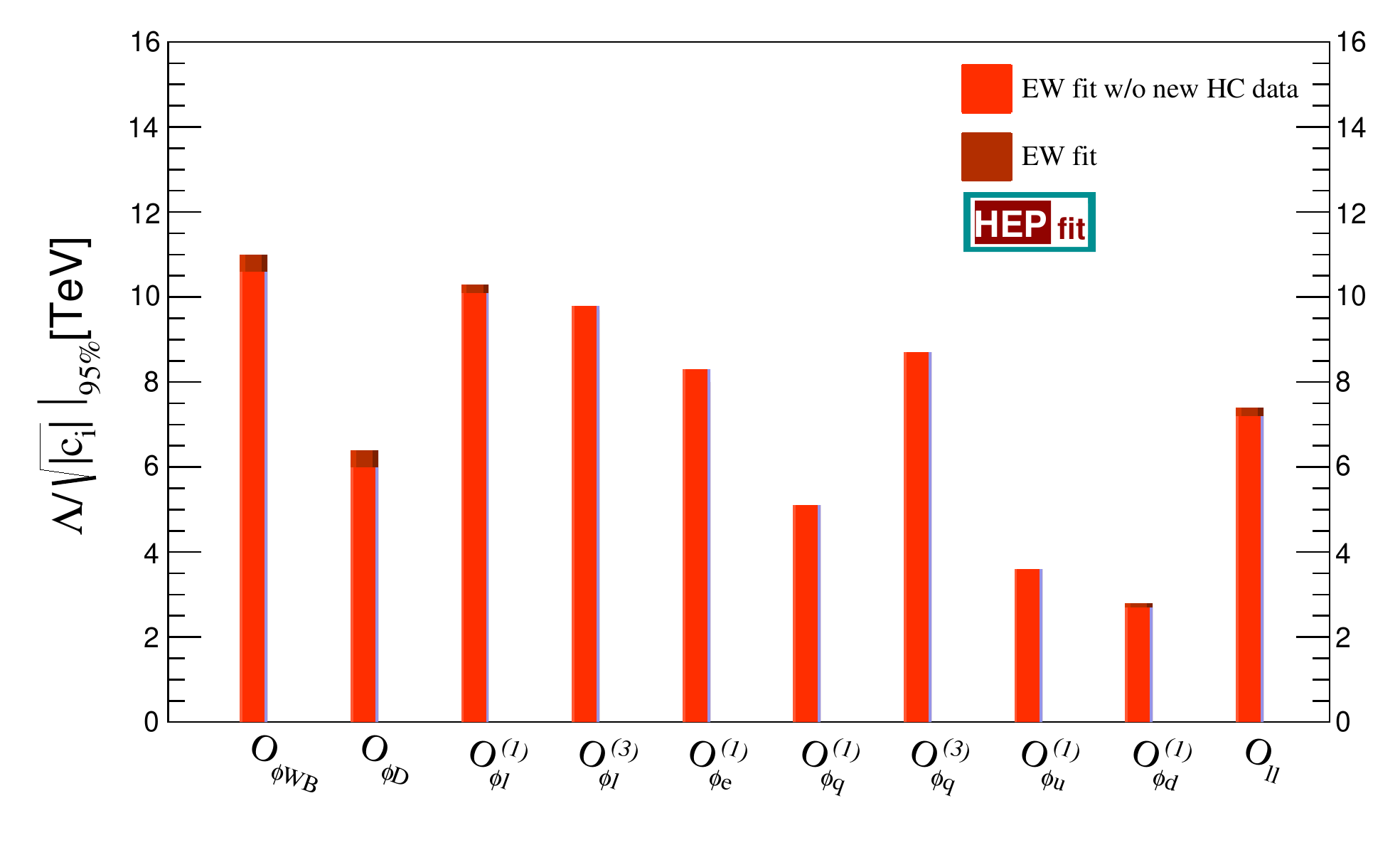} 
  \end{tabular}
  \vspace{-0.65cm}
  \caption{(Left) $68\%$ and $95\%$ probability limits on the dimension-6 operator coefficients $c_i/\Lambda^2$ [TeV$^{-2}$] from the fit to EWPO including all operators (in blue), compared with the bounds obtained assuming only one operator at a time (in red). (Right) $95\%$ probability limits on the NP interaction scale for the fits assuming only one operator at a time, showing also the effect of including the new HC data in each fit.}
  \label{fig:Dim6EWPO}
  \end{center}
\end{figure}

A model-independent description of indirect effects of NP (consistent with the SM symmetries and spectrum at low energies) is provided by the SM Effective Field Theory (SMEFT). The Lagrangian of the SMEFT extends the SM with higher-dimensional operators encoding the low-energy effects of the NP upon integrating out the high-energy degrees of freedom~\cite{delAguila:2000rc},
\begin{equation}
{\cal L}_{\mathrm{Eff}}={\cal L}_{\mathrm{SM}}+\sum_d \frac{1}{\Lambda^{d-4}}{\cal L}_d={\cal L}_{\mathrm{SM}}+{\cal L}_{5}+\sum_i \frac{c_i}{\Lambda^2}{\cal O}_{i}^{(6)}+\cdots .
\label{eq:Leff}
\end{equation}
The expansion in Eq.~(\ref{eq:Leff}) has been truncated at the dimension-6 level, which parameterizes the leading order NP effects in most observables in the electroweak sector. We use the basis of Ref.~\cite{Grzadkowski:2010es}, where we refer the reader for the definitions of the dimension-6 interactions.
The results of the global fit to EWPO are summarized in Figure \ref{fig:Dim6EWPO}. The left panel shows the bounds on the Wilson coefficients, $c_i/\Lambda^2$, from a fit including all the independent operators entering in the EWPO, compared to the bounds derived assuming that only one operator is present at a time.\footnote{While there are 10 operators in \cite{Grzadkowski:2010es} that enter in EWPO, the fit can only constrain 8 combinations. In our case, we take this into account by performing a small change of basis that trades the operators ${\cal O}_{\phi WB}$ and ${\cal O}_{\phi D}$ with 2 interactions that do not enter in EWPO (but correct Higgs observables).} 
(See also \cite{Ciuchini:2013pca} for related work.)
The results indicate the presence of a significant correlation between the contributions from different operators.
Hence, saturating the actual $95\%$ probability limits would require a significant fine tuning in the high energy theory in order to reproduce the observed correlations. In cases where such alignment is not present in the ultraviolet completion, the limits obtained turning on only one operator at a time may provide a more realistic order-of-magnitude estimate of the actual constraints on the NP interaction scale (see right panel of Figure \ref{fig:Dim6EWPO}). 


\section{Update on the Higgs boson constraints at the LHC Run 2}
\label{sec:Higgsfit}

\begin{figure}[t]
\begin{center}
  \begin{tabular}{c c c}
 \hspace{-1.5cm}\includegraphics[width=.33\textwidth]{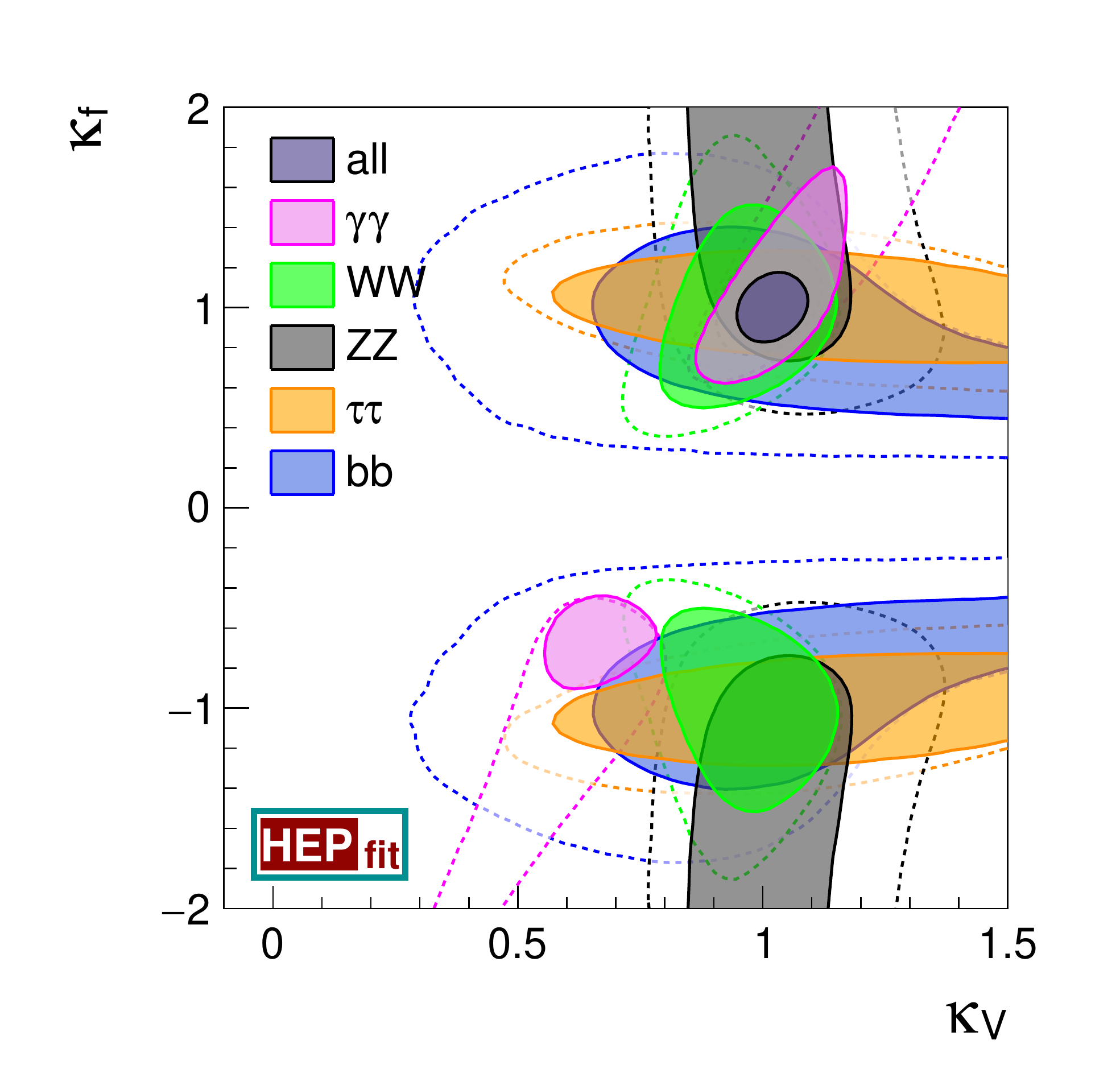} 
&
  \hspace{-0.8cm}\includegraphics[width=.33\textwidth]{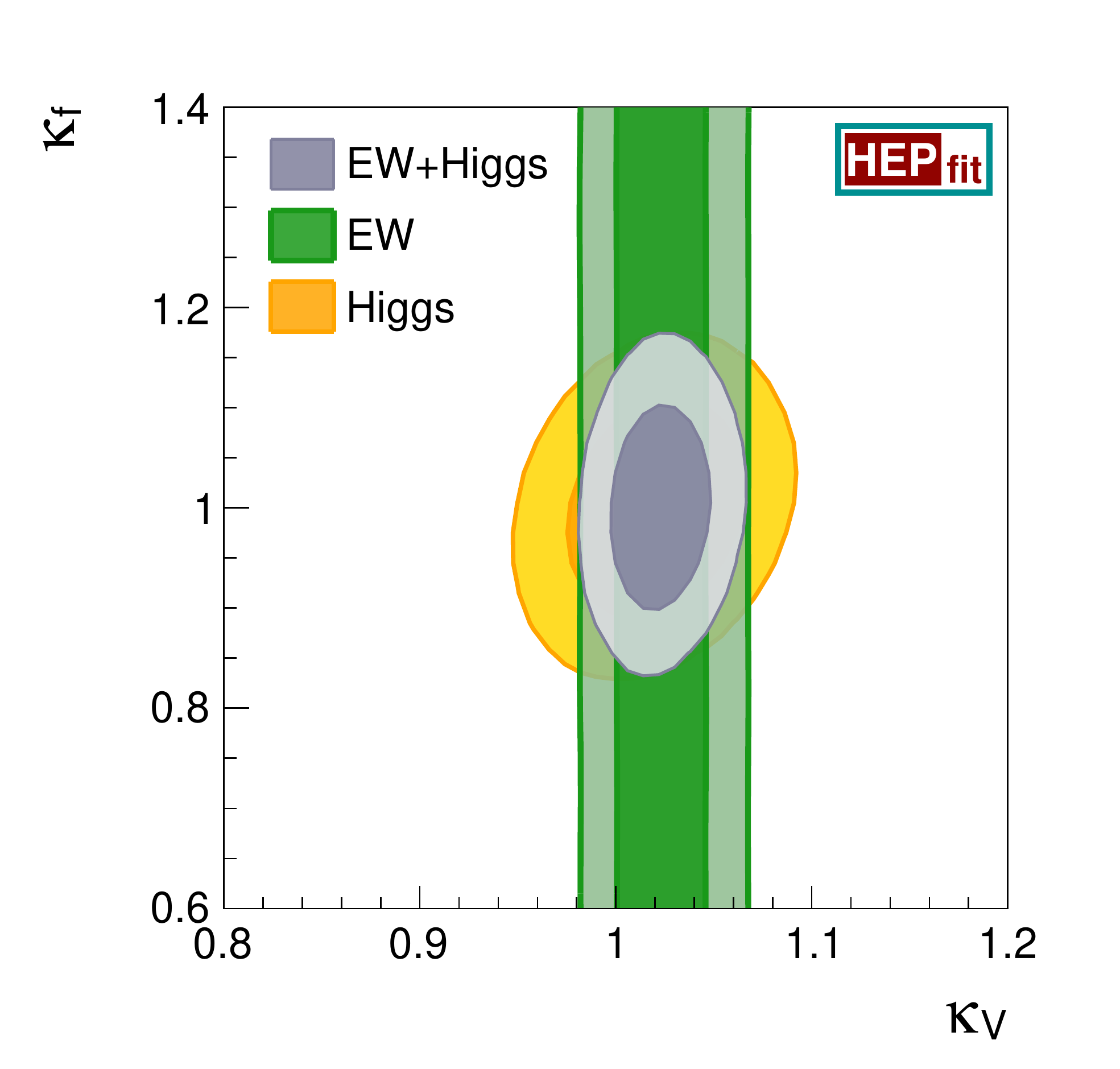} 
&
 \hspace{-0.8cm}\includegraphics[width=.5\textwidth]{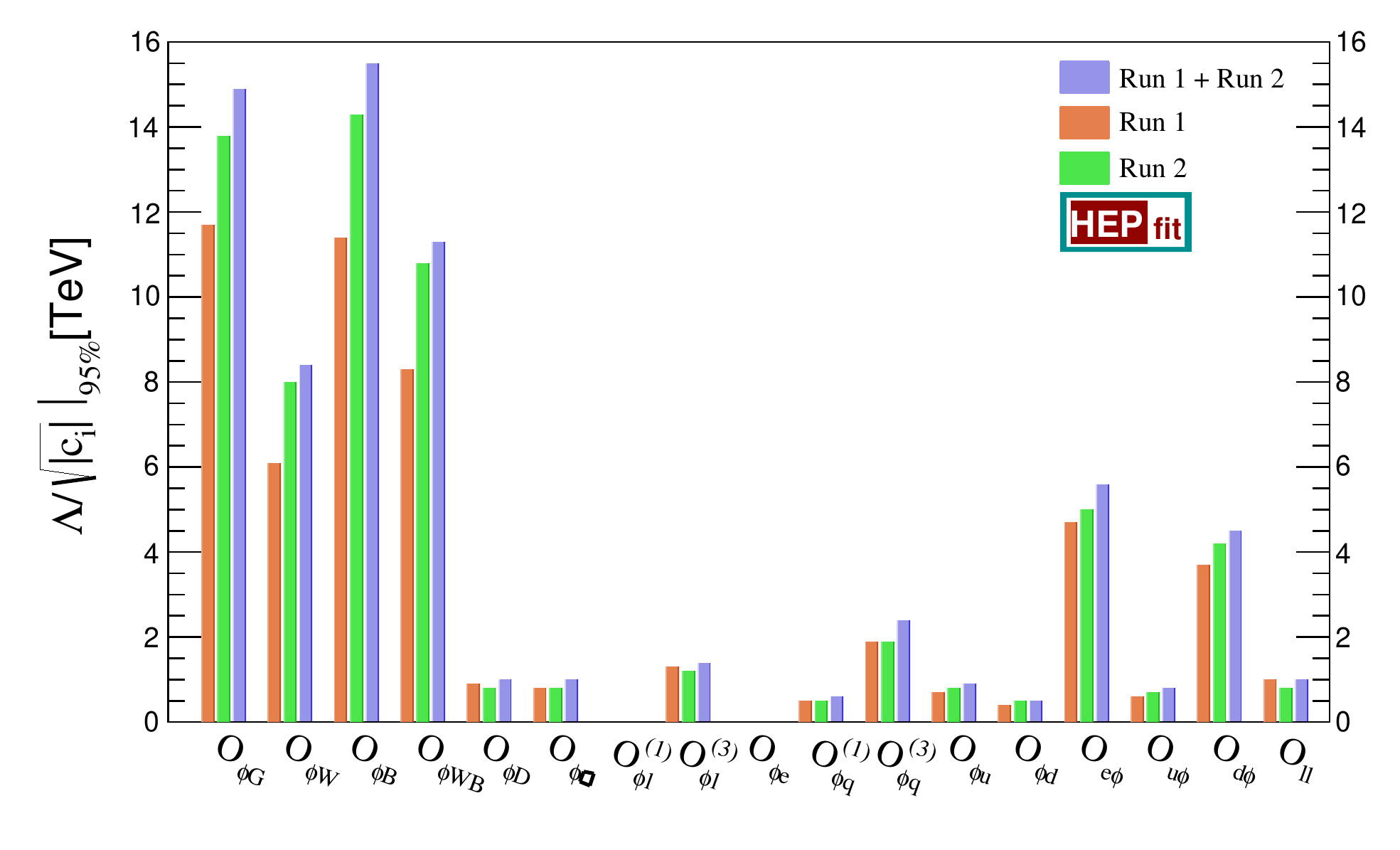}
  \end{tabular}
  \vspace{-0.65cm}
  \caption{(Left) Comparison of the 95$\%$ probability contours on the $\kappa_V$-$\kappa_f$ plane allowed by each Higgs decay channel using Run 1 (dashed lines) and Run 1+2 data (solid regions).  (Center)
  Comparison of the 68$\%$ and 95$\%$ probability contours in the same plane, from EWPO and current Higgs signal strengths (see \cite{deBlas:2016ojx} for details). 
  (Right) 95$\%$ probability limits on the NP interaction scale from the fit to each dimension 6 operator in the SMEFT (1 operator at a time).}
  \label{fig:HiggsLim}
  \end{center}
\end{figure}

In this section we discuss the impact of the latest measurements of the Higgs boson signal strengths at the LHC Run 2~\footnote{Including all data as of September 2017. See \cite{deBlas:2014ula} for previous results using only Run-1 data.} in constraining NP beyond the SM. 
For illustration purposes, in the left panel of Figure~\ref{fig:HiggsLim} we show the improvements obtained with Run-2 data in the $\kappa_V$-$\kappa_f$ plane for the different Higgs decay channels, with $\kappa_V$ ($\kappa_f$) a universal rescaling of the Higgs boson couplings to vector bosons (fermions).
When combined, despite the improvement in the constraints, we observe that the bounds on $\kappa_V$ are still dominated by the indirect effects in the EWPO (see central panel in Figure~\ref{fig:HiggsLim}). 

Turning our attention back to the dimension-6 SMEFT,
the right panel of Figure~\ref{fig:HiggsLim} shows the results from the fits to the interactions entering in Higgs observables, assuming one operator at a time. With $\sim36$ fb$^{-1}$ the effect of the 13 TeV results are already starting to dominate the bounds on several of the dimension-6 operators. Also, comparing Figures~\ref{fig:Dim6EWPO} and \ref{fig:HiggsLim}, we see that, with the exception of the operator ${\cal O}_{\phi WB}$ the limits from EWPO and Higgs observables are complementary on the dimension-6 parameter space. 
The results of a global fit including all operators simultaneously are however more intricate. There are again large correlations between the different NP effects, and somewhat flat directions allowing some of the interactions to go beyond the regime of validity of perturbation theory. In such cases there is a strong sensitivity to the effect of quadratic terms from the dimension-6 operators in the amplitudes squared. This can help to bound more efficiently the different operators, at the expense of limiting the range of applicability of the EFT results. 
The discussion of the results of a complete global fit will be provided elsewhere.


\section{Conclusions}

In these proceedings we have presented a preliminary study of the effects that the electroweak precision measurements taken at the Tevatron and LHC have on the global electroweak fit. While improvements in the electroweak precision constraints on NP are minor, it is remarkable that the recent hadron collider measurements of $\sin^2{\theta_{\mathrm{eff}}^{\mathrm{lept}}}$ are already competing in precision with the results from LEP and SLD. Further improvements are also expected in the determination of the $W$ mass, both from the full Tevatron data set as well as with future measurements from ATLAS and CMS. These could bring the overall precision close to the current theoretical uncertainty, allowing to test the SM prediction to a new level of accuracy.

We have also studied in these proceedings the Higgs-boson observable constraints obtained using the LHC 13 TeV data, and shown quantitatively the improvements already obtained compared with the Run-1 data. A more detailed study of these results will be presented in a future publication.


\section*{Acknowledgments}

The research leading to these results has received funding from the European Research Council under the European Union's Seventh Framework Programme (FP/2007- 2013) / grant n. 279972.



 
\end{document}